\documentclass[aps,pra,twocolumn,showpacs,superscriptaddress,floatfix,10pt]{revtex4}
\usepackage{graphicx}
\usepackage{psfrag}
\usepackage{amssymb,amsmath}
\usepackage{physics}
\usepackage{float}
\begin{document}
\newcommand{\beq}{\begin{equation}}
\newcommand{\eeq}{\end{equation}}
\newcommand{\ben}{\begin{eqnarray}}
\newcommand{\een}{\end{eqnarray}}
\newcommand{\bea}{\begin{array}}
\newcommand{\eea}{\end{array}}
\newcommand{\om}{(\omega )}
\newcommand{\bef}{\begin{figure}}
\newcommand{\eef}{\end{figure}}
\newcommand{\leg}[1]{\caption{\protect\rm{\protect\footnotesize{#1}}}}
\newcommand{\ew}[1]{\langle{#1}\rangle}
\newcommand{\be}[1]{\mid\!{#1}\!\mid}
\newcommand{\no}{\nonumber}
\newcommand{\etal}{{\em et~al }}
\newcommand{\geff}{g_{\mbox{\it{\scriptsize{eff}}}}}
\newcommand{\da}[1]{{#1}^\dagger}
\newcommand{\cf}{{\it cf.\/}\ }
\newcommand{\ie}{{\it i.e.\/}\ }   

\newcommand{\spazio}{\vspace{0.3cm}}
\hyphenation{bio-mol-ecules}
\newcommand{\de}[1]{\frac{\partial}{\partial{#1}}}
\newcommand{\U}{\tilde{U}}
\newcommand{\V}{\tilde{V}}

\title{Optimal Encoding Capacity of a Linear Optical Quantum Channel}

\author{J.~A.~Smith}
\affiliation{Tulane University, Department of Physics, New Orleans, Louisiana 70118, USA}
\author{D.~B.~Uskov}
\affiliation{Tulane University, Department of Physics, New Orleans, Louisiana 70118, USA}
\affiliation{Brescia University, Department of Mathematics and Natural Science, Owensboro, Kentucky 43201, USA}
\author{L.~Kaplan}
\affiliation{Tulane University, Department of Physics, New Orleans, Louisiana 70118, USA}

 \begin{abstract}
 	Here, we study the classical information capacity of a quantum channel, assuming linear optical encoding, as a function of available photons and optical modes. We present a formula for general channel capacity and show that this capacity is achieved without requiring the use of entangling operations typically required for scalable universal quantum computation, e.g. KLM measurement-assisted transformations. As an example, we provide an explicit encoding scheme using the resources required of standard dense coding using two dual-rail qubits (2 photons in 4 modes). In this case, our protocol encodes one additional bit of information. Greater gains are expected for larger systems.
\end{abstract}                                                               
                                                                            
\date{\today}
\maketitle

\label{sec_model}
\section{Introduction}
Superdense coding is an elegant application of basic quantum mechanics that can provide exciting gains in the capacity of an information channel. A simple protocol proposed in 1992 by Bennett and Wiesner proved that two classical bits of information could be sent over a quantum channel via a single qubit~\cite{Bennett}, a compression made possible by exploiting entanglement as a physical resource. Entanglement can manifest in many forms and in a variety of physical hosts; engineering an entangled system to be used as an efficient communication channel thus presents an interesting challenge. In general, an optimal quantum channel will require a well-devised protocol to properly manipulate a maximally entangled state~\cite{Horodecki}. 

We focus here on a linear optical~\cite{Kok} quantum communication channel constructed from a set of modes and photon quanta. The Hilbert space dimension for such a system of \textit{N} total photons propagating in \textit{M} optical modes is given by
\begin{equation}
	d_H = \frac{(N+M-1)!}{N!(M-1)!} \,,
\end{equation}
or the number of ways to place $N$ photons in $M$ modes.
Considering, for example, a system of two dual-rail qubits ($N=2$, $M=4$), we note that $d_H=10$ whereas the logical qubit space is only four-dimensional. Resource efficiency is paramount in communication and such overhead immediately limits channel capacity. 
In this paper, we aim to develop optimal dense coding protocols specifically for a linear optical quantum channel, without imposing the qubit as an operational basis.

\begin{figure}[ht]
	\spazio
	\centering
	\includegraphics[width=0.5 \textwidth]{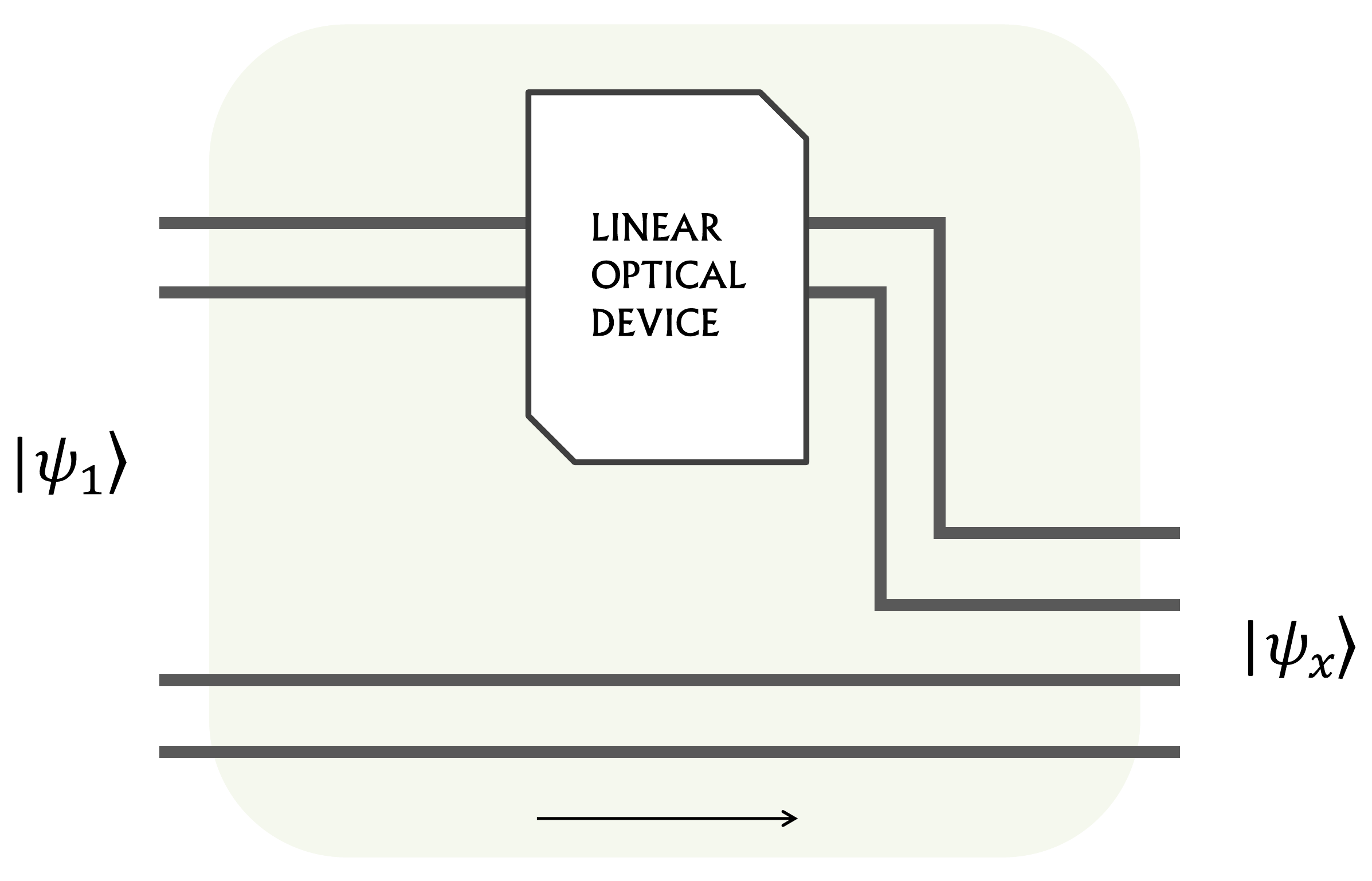}
	\caption{(Color online) A typical quantum optical encoding device consisting of 4 modes. 2 modes have been distributed to Alice (top), and 2 modes have been distributed to Bob (bottom). Alice applies a linear optical operation and sends her resources to Bob who performs a measurement on the total state.}
	\label{Apparatus}
\end{figure}

The protocol requires that Alice and Bob initially share an entangled state of $N$ photons, $\ket{\psi_1}$, with $M_A$ out of the total $M$ modes under Alice's control. Alice chooses a classical symbol $x$ and encodes it by operating on her modes, taking $\ket{\psi_1}$ to $\ket{\psi_x}$. Her resources are forwarded to Bob, who then performs a quantum measurement on the entire state and reads out a classical bit string. Refer to Fig.~(\ref{Apparatus}).
$d_H$ serves as an  absolute upper bound on the number of distinguishable states Alice can send, though due to the restriction of utilizing only linear optical operations (beam splitters and phase shifters), Alice may not be able to make use of the entire Hilbert space, even if she possesses all of the modes ($M_A=M$). The crux of the problem is that not all quantum logic operations are guaranteed to be accessible via linear optical transformations. Entangling operations in particular require photons to interact, a task they are notoriously ill-suited for. Solutions to this problem exist, e.g. the use of KLM~\cite{KLM} auxiliary resources or non-linear Kerr media. For now, we omit such components from our system. As we shall determine, linear optics is sufficient to construct a maximally efficient communication channel.

We represent the state of our system in the Fock basis
\begin{equation}
\ket{n_1 n_2... n_M} = \frac{(\hat{a}_1^\dagger)^{n_1}(\hat{a}_2^\dagger)^{n_2}...(\hat{a}_M^\dagger)^{n_M}}{\sqrt{n_1!n_2!...n_M!}}\ket{0} \,.
\end{equation}
 A general linear operation can then be expressed as
 \begin{equation}
\hat{a}_i^\dagger \rightarrow \sum_{j=1}^{M_A} u_{ij}\hat{a}_j^\dagger \quad \forall \;\; i =1 \ldots  M_A \,,
\label{lintransf}
\end{equation}
where $u_{ij}$ are the matrix elements of a unitary operation $U$ acting on $M_A$ 
modes. Alice encodes one of $\abs{X}$ classical symbols into the system by generating the appropriate quantum state, $\ket{\psi_1},\ket{\psi_2},\ldots,\ket{\psi_{\abs{X}}}$. Each state $\ket{\psi_x}$ is constructed by applying a different transformation $U_x$ to the initial entangled state, $\ket{\psi_1}$ (here without loss of generality we will take $U_1=I$). 

The mutual entropy $H(X:Y)$ shared between Alice and Bob is constrained by the Holevo bound, which for pure states can be written as~\cite{Holevo,Hausladen}
\begin{equation}
	H(X:Y) \le_1 S(\rho) \le_2 H(X) \le_3 \log_2 \abs{X},
	\label{HolevoB}
\end{equation}
where $X$ and $Y$ are the encoding and decoding variables,
\begin{equation}
	\rho=\sum_{x=1}^{\abs{X}} p_x \ket{\psi_x}\bra{\psi_x}, 
\end{equation}
$p_x$ is the probability of choosing symbol $x$, $S(\rho)$ is the von Neumann entropy of $\rho$, and $H(X)$ is the Shannon entropy of Alice's encoding variable.
Inequality  1 reflects the fact that depending on Bob's measurement capabilities, he may not be able to access all the information encoded by Alice; it becomes an equality only if Bob can perform arbitrary measurements on words of arbitrary length~\cite{Hausladen,Giovannetti,Lloyd}. We return to this point in Sec. VI, and in the following we focus on $S(\rho)$, or the encoding capacity of the channel. Inequality 2 becomes an equality when all contributions $\ket{\psi_x}$ to $\rho$ are orthogonal, and finally inequality 3 becomes an equality when all symbols in $X$ are chosen with equal probability. 

\section{An Upper Bound on Encoding Capacity}
From Eq.~(\ref{HolevoB}), it immediately follows that encoding capacity is limited by the Hilbert space dimension,
\begin{equation}
	S(\rho) \le \log_2(d_H).
	\label{UltUppBound}
\end{equation}
 This inequality provides a valuable reference point, but for our purposes we can derive an even stricter bound on $S(\rho)$. First, we abstractly define $d_S$ as the dimension of the span of the states Alice can generate via linear optical transformations as presented in Eq.~(\ref{lintransf}). The Holevo bound again restricts the encoding capacity,
 \begin{equation}
 	S(\rho) \le \log_2(d_S) \quad \quad d_S \le d_H \,.
 	\label{spanUpperBound}
 \end{equation}
Then, we note the Hilbert space of our channel can be decomposed into a direct sum
\begin{equation}
\mathcal{H} = \bigoplus_{N_A=0}^{N} (\mathcal{H}^{A,N_A} \otimes \mathcal{H}^{B,N_B})\,,
\label{subspacebreakdown}
\end{equation}
where each term represents the subspace corresponding to a fixed distribution of photons between Alice and Bob. $N_A$ is the number of photons in Alice's modes, and $N_B=N-N_A$. Any initial state $\ket{\psi_1}$ can be decomposed into a sum of components 
\begin{equation}
\ket{\psi_{N_A}} \in \mathcal{H}^{A,N_A} \otimes \mathcal{H}^{B,N_B} \,.
\end{equation}
For each such component we can construct a Schmidt decomposition over a set of $\min(d_H^{A,N_A},d_H^{B,N_B})$ basis vectors, where $d_H^{A,N_A}$ and $d_H^{B,N_B}$ are the dimensions of $\mathcal{H}^{A,N_A}$ and $\mathcal{H}^{B,N_B}$ respectively. \textit{Any} operation on Alice's modes will act only on the basis vectors of her subspace. 
It follows that for any $\ket{\psi_{N_A}}$ where $d_H^{A,N_A} < d_H^{B,N_B}$, Alice has no control over the dimensions of Bob's subspace which are not spanned by the basis vectors of the Schmidt decomposition. Thus, the upper bound for the dimension of the span of the set of states Alice is able to generate within each term of Eq.~(\ref{subspacebreakdown}) is given by
\begin{equation}
d_S^{N_A} \le d_H^{A,N_A} \min(d_H^{A,N_A},d_H^{B,N_B})\,.
\label{termupperbound}
\end{equation}
We combine Eqs.~(\ref{subspacebreakdown}) and (\ref{termupperbound}) to find the upper bound on the total span $d_S$ for given photon number $N$ and modes $M_A, M_B \ge 1$:
\begin{equation}
d_S \le \sum_{N_A=0}^{N} f(N_A)\,,
\label{mainresult}
\end{equation}
where 
\begin{equation}
\small f(N_A)= g(N_A,M_A) \min(g(N_A,M_A),g(N_B,M_B)) \,,
\label{f}
\end{equation}
\begin{equation}
g(n,m)=\frac{(n+m-1)!}{n!(m-1)!} 
\end{equation}
is the number of ways to distribute $n$ photons over $m$ modes,
and $M_B=M-M_A$.

This is a key result; if we combine Eqs.~(\ref{spanUpperBound}) and (\ref{mainresult}) we find an upper bound on the information capacity for a given physical device. As we will see in Sec. III, we find that the bound obtained here is tight.

\section{Numerical Testing of Alice's Encoding Capabilities}
Eqs.~(\ref{spanUpperBound}) and (\ref{mainresult}) provide a fully analytic upper bound for the information capacity of an optical channel, but the question remains whether and how this bound can actually be reached using linear optical encoding. To obtain the actual encoding capacity, we numerically maximize $S(\rho)$ over the initial state $\ket{\psi_1}$ and Alice's transformations $U_x$. We call the result of this optimization $S_{max}$.
Similarly, $d_S$ is computed by evaluating the rank of the density matrix for an ensemble of states generated with a sufficiently large set of random $U_x$.

\begin{figure}[ht]
 	\spazio
 	\centering
 	\includegraphics[width=0.25 \textwidth,height=2.5in]{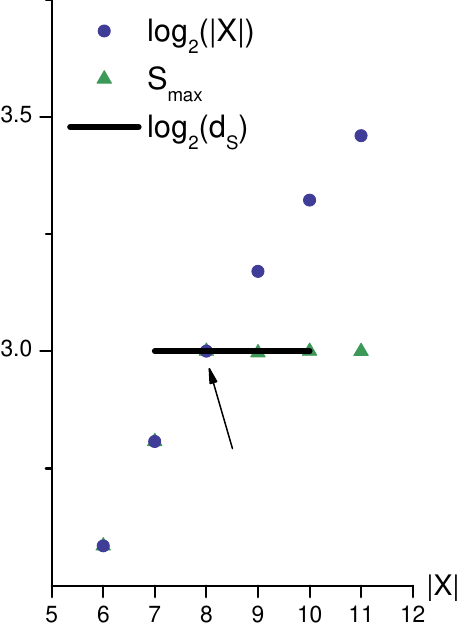}\includegraphics[width=0.25 \textwidth,height=2.5in]{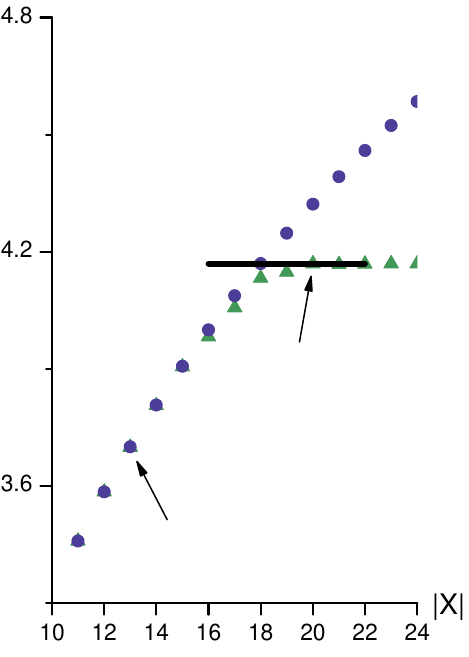}
 	\caption{(Color online) Maximization of the von Neumann entropy $S(\rho)$ as a function of the number of symbols $\abs{X}$. (Left) $N=2$, $M=4$, $M_A=2$. A clear convergence to $\log_2(8)=3$ is observed. (Right) $N=3$, $M=5$, $M_A=2$. Here $S_{max}$ reaches $\log_2 \abs{X}$ for $\abs{X} \le 13$, indicating that up to $13$ orthogonal states may be generated. Globally, $S_{max}$ reaches $\log_2(18)$ at $\abs{X}=20$. Analytic solutions can be constructed for $\abs{X} \le 13$.}
 	\label{fig_optim}
\end{figure}

For small $N$ and $M$, the numerical results are fairly straightforward and elegant. Of particular interest is the case where $N=2$, $M=4$, $M_A=2$, 
which uses the same physical resources as the standard, two dual-rail qubit dense coding protocol. Here we find that $d_S=8$, in agreement with the right hand side of Eq.~(\ref{mainresult}). Furthermore, we find that $S_{max} \rightarrow \log_2 (d_S) = \log_2(8)$ bits can be achieved by generating $|X|=8$ completely distinguishable states, see Fig.~\ref{fig_optim} (Left).
In contrast with the standard dense coding protocol, which allows Alice to communicate 2 classical bits via a single photon in two modes, we instead find that Alice is able to send 3 classical bits via 1.25 photons (2.4 bits per photon or 1.5 bits per optical mode). This is an important result; Alice is able to encode additional information into the channel by not restricting herself to the qubit basis. The advantage gained by expanding the optimization space beyond the dual rail basis has been consistently observed in other work on gate optimization~\cite{Uskov} and photonic quantum communication~\cite{LU}.

For larger, more resource-intensive systems, such straightforward convergence of $S_{max}$ to $\log_2 \abs{X}$ is not observed. We instead find an intermediate optimization regime where the encoding capacity is achieved by increasing the number of coded states $\abs{X}$ beyond the number of orthogonal states Alice can send. As an example, we present the results for $N=3$, $M=5$, $M_A=2$ in Fig.~\ref{fig_optim} (Right).
Here, we find $d_s=18$ [again in agreement with the right hand side of Eq.~(\ref{mainresult})] and indeed $S_{max} \rightarrow \log_2(d_S) = \log_2(18)$, but only when $\abs{X} \ge 20$. The maximum channel capacity is equivalent to one associated with 18 orthogonal quantum states, however Alice cannot generate 18 such states using linear optics. Instead, she must generate a set of at least 20 non-orthogonal states to obtain the maximum capacity of the channel. It becomes apparent that Alice cannot, in general, construct a set of $d_S$ orthonormal basis vectors over the entire subspace accessible to her via allowed optical transformations. She must instead saturate the system with additional linearly independent states until she reaches the potential of the hardware. 

\begin{figure}[ht]
	\spazio
	\centering
	\includegraphics[width=0.5 \textwidth]{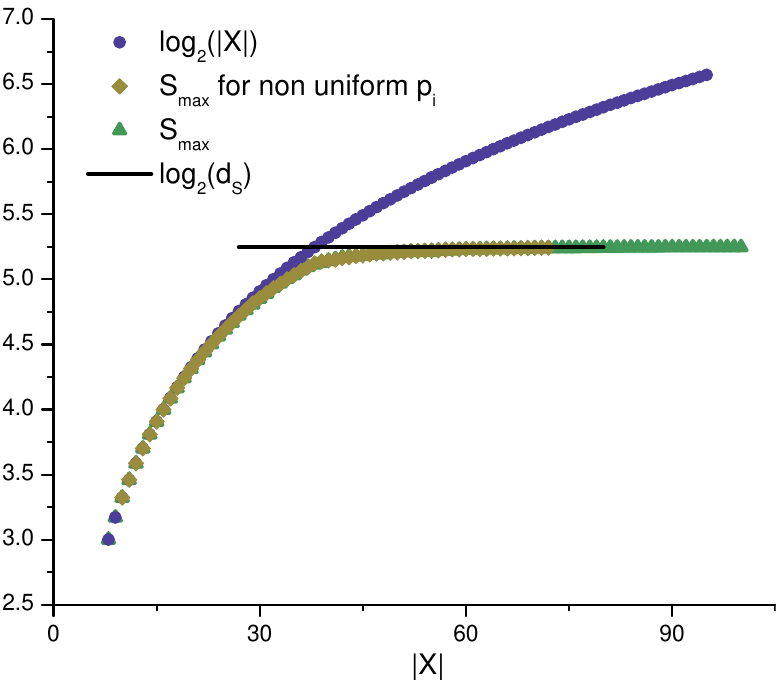}
	\caption{(Color online) Maximization of the von Neumann entropy for $N=3$, $M=6$, $M_A=3$. $\abs{X}$ must be increased well beyond $d_S=38$ for the von Neumann entropy $S_{max}$ to maximize to $\log_2(38)$. No gain in information capacity is observed if we allow Alice to send symbols from $X$ with non-uniform probability.}
	\label{fig363}
\end{figure}

In  Fig.~\ref{fig363}, we present a larger system,  $N=3$, $M=6$, $M_A=3$, where $d_S= 38$. Here the full encoding capacity $\log_2 d_S$ is approached slowly as the number of symbols increases, although a modest number of symbols is sufficient to attain a very large fraction of the maximum capacity.

Empirically, we observe in all cases that $d_S$ and $S_{max}$ do attain the upper bounds defined in Eq.~(\ref{mainresult}) and Eq.~(\ref{spanUpperBound}), respectively, assuming only linear optical encoding. We therefore define classical information capacity $C$ in bits per channel use as a function of $N, M, M_A$:
\begin{equation}
	C(N,M,M_A) = \log_2(d_S) = \log_2\left (\sum_{N_A=0}^{N} f(N_A)\right)\,,
	\label{supermainresult}
\end{equation}
where $f(N_A)$ is defined in Eq.~(\ref{f}). Because the bound in Eq.~(\ref{mainresult}) holds for \textit{any} photon-number preserving operation on Alice's modes, the use of non-deterministic or non-linear entangling optical components will not improve channel capacity.

\section{Asymptotic Behavior of the Encoding Capacity}
To gain some insight into the parameter dependence implied by  Eq.~(\ref{supermainresult}), we consider the thermodynamic limit $M= \alpha N$, for $M_A,N_A,M_B,N_B,M,N \gg 1$. Applying Stirling's approximation, the full Hilbert space dimension is given by
\begin{equation}
\log_2 d_H \simeq N [\log_2(1+\alpha)+\alpha \log_2\left(1+\alpha^{-1}\right)]\,.
\end{equation}
Similarly, 
\begin{equation}
   \log_2 g(n,m) \simeq m \log_2(1+\frac{n}{m})+ n \log_2(1+\frac{m}{n})\,,
   \label{Stirlings Approximation g}
\end{equation}
and thus $N_A^{p}$, the maximum of $g(N_A,M_A) g(N_B,M_B)$, is given by
\begin{equation}
	N_A^{p} \simeq N M_A/M\,.
	\label{h_peak}
\end{equation}
Now $f(N_A)$ is piecewise defined with the crossover point $N_A^{c}$ determined by $M_A$. 
One can verify that $N_A^{p} > N_A^{c}$ if $\frac{M_A}{M_B}>1$, $N_A^{p} = N_A^{c}$ if $\frac{M_A}{M_B}=1$, and $N_A^{p} < N_A^{c}$ if $\frac{M_A}{M_B} < 1$. Thus, the value of $f(N_A)_{max}$ is dependent on the ratio of modes $M_A/M_B$. We find the following regimes of behavior as a function of the mode ratio $M_A/M_B$: 
\begin{eqnarray}
 & & \frac{M_A}{M_B}>1 \Rightarrow N_A^{p} > N_A^{c} \Rightarrow d_S \simeq d_H \label{Regime 1} \\
 & & \frac{M_A}{M_B}=1 \Rightarrow N_A^{p}=N_A^{c}=\frac{N}{2} \Rightarrow d_S \simeq \frac{d_H}{2} \label{Regime 2} \\
 & & \frac{M_A}{M_B}<1 \Rightarrow N_A^{p} < N_A^{c} \Rightarrow d_S \ll d_H \\
 & & M_B \ge g(N-1,M_A) \nonumber \\ & & \Rightarrow d_S \simeq g^2(N-1,M_A) \ll d_H \label{Regime 4}
\end{eqnarray}
In particular, we see from Eq.~(\ref{Regime 1}) that Alice maintains near absolute control of the channel in the range $M_A/M_B>1$, and the maximum encoding capacity is asymptotically indistinguishable from the capacity implied by the total Hilbert space dimension $d_H$. Eq.~(\ref{Regime 2}) indicates that when the modes are evenly split between Alice and Bob, the maximum encoding capacity is asymptotically lower by precisely one bit than the capacity implied by $d_S$.  From Eq.~(\ref{Regime 4}), we find that for a fixed number of modes $M_A$ controlled by Alice, any modes Bob possesses in excess of $g(N-1,M_A)$ do not contribute to channel capacity.
\begin{figure}[H]
	\spazio
	\centering
	\includegraphics[width=0.45 \textwidth]{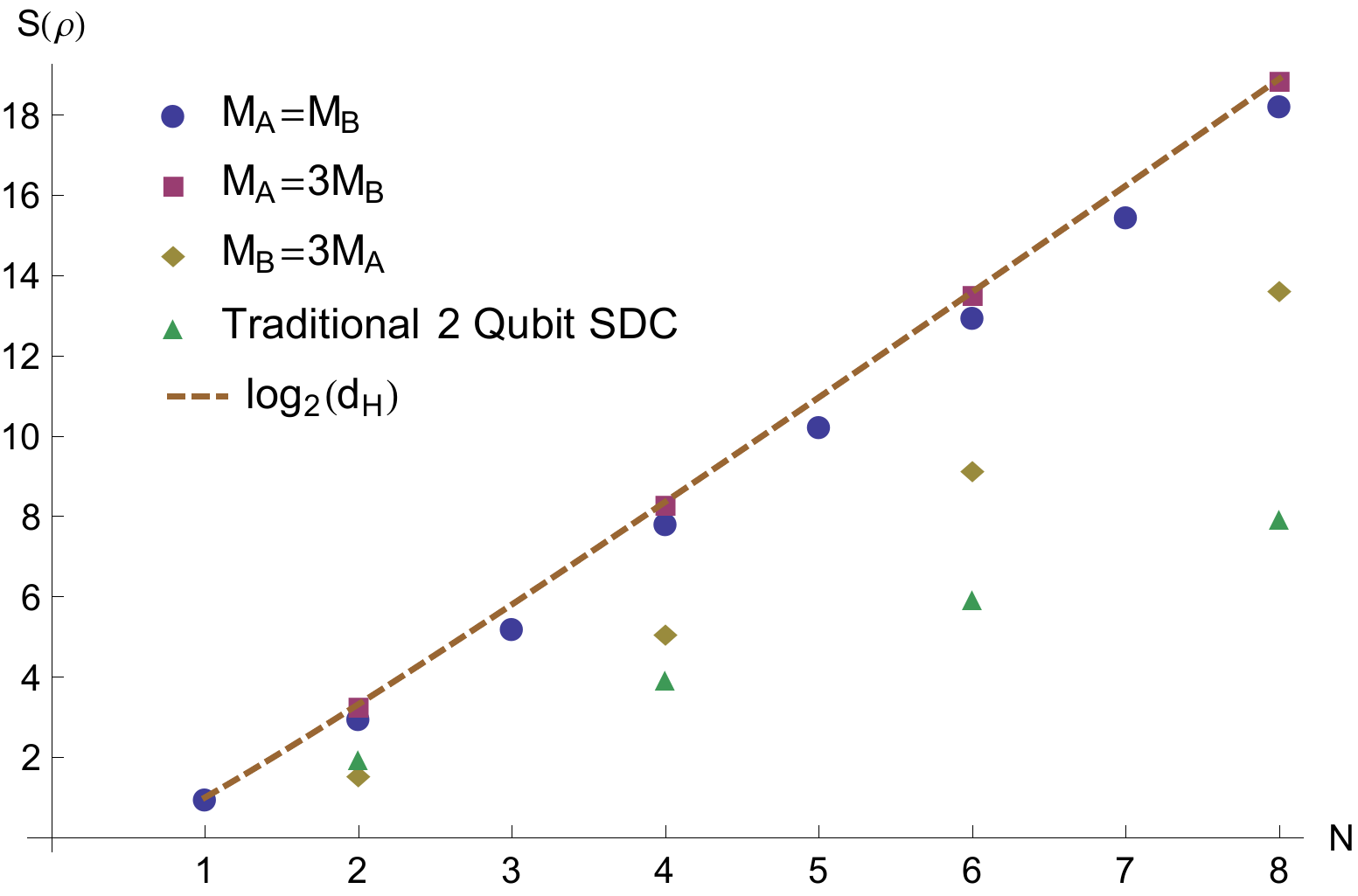}
	\caption{(Color online) Encoding capacity $\log_2 d_S$ for $M=2N$ and different ratios $M_A/M_B$. As an upper and lower bound, we include, respectively, the Hilbert space dimension and the capacity of traditional dense coding using two dual-rail qubits (which assumes $M_A=M_B$).}
	\label{potential}
\end{figure}
\section{Developing Explicit Encoding Protocols}
Thus far we have presented a general method for calculating the information capacity of linear optical hardware. An important task remains; we must be able to design a dense coding protocol for actual implementation. We return to the prototypical case: $N=2$, $M=4$, $M_A=2$. Using the results of the numerical optimization of $S(\rho)$, we find the general set of initial states and transformations that satisfy $\braket{\psi_i}{\psi_j}=0$  $\forall$  $i \ne j$. 
The result is a non-trivial set of equations, which simplify elegantly if we slightly restrict the space of Alice's transformations, $U_3, U_4, \ldots, U_8$, see Eq.~(\ref{DCexample}).

\section{Discussion}
We have already noted that the Holevo bound is not automatically reachable, and the information capacity of an optical quantum channel is limited by factors other than efficient encoding. The susceptibility of entangled states to decoherence and noise must be considered. Optical quantum computers are an attractive idea in this regard because photons do not strongly interact with most matter; photonic states can thus be carried over large distances~\cite{KLM, Uskov}.

So far we have not addressed the practicality of generating the pre-distributed maximally entangled initial states. The question remains as to whether $\ket{\psi_1}$ can be constructed using deterministic optical components. This will likely depend on the physical resources ($N,M,M_A$) of a particular device. While deterministic state generation is certainly preferred, it is not essential for efficient communication if one considers cost as that associated with channel usage. 

At the receiving end of a quantum channel, signal decoding is an especially challenging task. For the case in which Alice sends orthogonal states, the Holevo bound is achieved by applying an appropriate von Neumann measurement. In principle, a sufficiently large subset of von Neumann measurements can be carried out by having Bob rotate incoming states into the photo-counting basis states (see Lougovski and Uskov~\cite{LU}). We find that it is not possible to implement such a measurement using deterministic linear optics exclusively. If one were to \textit{impose} a linear optical measurement, Bob would be unable to fully distinguish incoming states~\cite{Kwiat}. To make things even more complicated, our results suggest that in order to optimize $S(\rho)$, Alice will often send a set of non-orthogonal states. Thus, even if we were able to apply an optimal POVM, $H(X:Y)$ would fall well below $S(\rho)$ for a single communication~\cite{Holevo,Peres}. Achieving the Holevo bound for an \textit{a priori} set of non-distinguishable quantum states is a common challenge for quantum information theory, and has been the subject of some study. Proposed solutions~\cite{Hausladen,Giovannetti,Lloyd,Ogawa} rely on sending code words comprised of multiple states, $\ket{\psi_a \psi_b ...}$. For example, the PGM~\cite{Hausladen} scheme guarantees that a word composed of $l$ states carries $H(X:Y) = l S(\rho)$ information for a coding vocabulary in the limit of large $l$. The caveat is that Bob must store previously sent letters in a quantum memory storage device, so that he may act on an entire word with a joint measurement. This could be implemented using $l M$ storage modes and a rail-switching junction for incoming letters.
\begin{widetext}
\begin{equation}
	\begin{split}
	& \mbox{Initial entangled state: }\\
	& \begin{align*} \quad \ket{\psi_1}=c_1 e^{i d_1}\ket{2000}+c_2 e^{i d_2} \ket{1100}+c_3 e^{i d_3} \ket{1010}+c_4 e^{i d_4} \ket{1001}+c_5 e^{i d_5} \ket{0200} \end{align*} \\
	& \begin{align*} \quad +c_6 e^{i d_6} \ket{0110}+c_7 e^{i d_7} \ket{0101}+c_8 e^{i d_8} \ket{0020} +c_9 e^{i d_9} \ket{0011}+c_{10} e^{i d_{10}} \ket{0002}\,. \end{align*}\\
	& \\
	& \mbox{Alice's operations:}\\
	& \begin{align*}
	 U_1 = \begin{pmatrix} 1 & 0 \\ 0 & 1 \end{pmatrix}   \;\;\;\; U_2 = \begin{pmatrix} -1 & 0 \\ 0 & -1 \end{pmatrix}
	\end{align*} \\
	& \begin{align}
	U_x = \begin{pmatrix} i(-1)^x  \sqrt{\frac{1}{3}} & -\sqrt{\frac{2}{3}}e^{-i q_x} \\ \sqrt{\frac{2}{3}}e^{i q_x} & i(-1)^{x+1}  \sqrt{\frac{1}{3}} \end{pmatrix} \quad (x=3,4, \ldots,8)
	\end{align} \\
	& \begin{align*}
	q_x=q_3+\frac{\pi}{3}(x-3) \;\;\;\; (x=4,\ldots,8)
	\end{align*} \\
	& \mbox{Restricted by:}\\
	& \begin{align*} c_1^2+c_2^2+c_5^2=\frac{3}{8} \quad c_3=c_7 \quad c_4=c_6 \quad c_3^2+c_4^2=\frac{1}{4} \quad c_8^2+c_9^2+c_{10}^2=\frac{1}{8} \end{align*} \\
	& \begin{align*} d_3+d_7-d_4-d_6= m \pi \quad m \in \mathbb{Z}_{odd} \end{align*} \\
	& \begin{align*} c_1 c_2 \cos(d_1-d_2-q_3)-c_1 c_5 \sin(d_1-d_5-2 q_3)-c_2 c_5 \cos(d_2-d_5-q_3)=0 \end{align*}\\
	& \begin{align*} c_1 c_2 \sin(d_1-d_2-q_3)-c_1 c_5 \cos(d_1-d_5-2 q_3)-c_2 c_5 \sin(d_2-d_5-q_3)=0 \end{align*} \\
	\end{split}
	\label{DCexample}
\end{equation}
\end{widetext}

\section{Conclusion}
In summary, we have shown that hardware-specific coding algorithms must be developed in order to make full use of a linear optical quantum channel. In Eq.~(\ref{supermainresult}), we obtained a simple expression for channel capacity as a function of photons and optical modes. Furthermore, we found this capacity is achievable via \textit{some} encoding procedure using linear optics exclusively. We observe that for small systems, Alice can send completely distinguishable states. Generally, however, Alice must choose from a pool of non-orthogonal states. In either case, it is possible to extract an encoding protocol from the numerical maximization of the von Neumann entropy. 
\newline
\newline
\begin{acknowledgments}
We are grateful for helpful discussions with Nick Sparks and Pavel Lougovski.
This work was supported in part by the NSF under Grants PHY-1005709 and PHY-1205788. DBU acknowledges support from AFRL Information Directorate under grant FA 8750-11-2-0218. This research was supported in part using high performance computing (HPC) resources and services provided by Technology Services at Tulane University.
\end{acknowledgments}


\begin{thebibliography}{99}

\bibitem{Bennett} C. H. Bennett and S. J. Wiesner, Phys. Rev. Lett. \textbf{69}, 2881 (1992).

\bibitem{Horodecki} R. Horodecki, P. Horodecki, M. Horodecki, and K. Horodecki, Rev. Mod. Phys. \textbf{81}, 865 (2009).

\bibitem{Kok} P. Kok, W. J. Munro, K. Nemoto, T. C. Ralph, J. P. Dowling, and G. J. Milburn, Rev. Mod. Phys. \textbf{79}, 135 (2007).

\bibitem{KLM} E. Knill, R. Laflamme, G. J. Milburn, Nature (London) \textbf{409}, 46 (2001).

\bibitem{Holevo} A. S. Holevo, Probl. Peredachi Inf. \textbf{9}, 110 (1973).

\bibitem{Hausladen} P. Hausladen, R. Jozsa, B. Schumacher, M. Westmoreland, and W. K. Wootters, Phys. Rev. A \textbf{54}, 1869 (1996).

\bibitem{Giovannetti} V. Giovannetti, S. Lloyd, and L. Maccone, Phys. Rev. A \textbf{85}, 012302 (2012).

\bibitem{Lloyd} S. Lloyd, V. Giovannetti, and L. Maccone, Phys. Rev. Lett. \textbf{106}, 250501 (2011).

\bibitem{Uskov} D. B. Uskov, A. M. Smith, and L. Kaplan, Phys. Rev. A \textbf{81}, 012303 (2010).

\bibitem{LU}  P. Lougovski and D. B. Uskov Phys. Rev. A \textbf{92}, 022303 (2015).

\bibitem{Kwiat} K. Mattle, H. Weinfurter, P.G. Kwiat, and A. Zeilinger, Phys. Rev. Lett. \textbf{76}, 4656 (1996).

\bibitem{Peres} A. Peres and W. K. Wootters, Phys. Rev. Lett. \textbf{66}, 1119 (1991).

\bibitem{Ogawa} T. Ogawa and H. Nagaoka, IEEE Trans. Inf. Theory \textbf{53}, 2261 (2007).

%
%
%
%
%
%
%
%
%
%


\end{thebibliography}
\end{document}